# Accurate wavelength tracking by exciton spin mixing

**Authors:** Anton Kirch[1], Toni Bärschneider[1], Tim Achenbach[1], Felix Fries[1], Max Gmelch[1], Robert Werberger[1], Chris Guhrenz[2], Aušra Tomkevičienė[1,3], Johannes Benduhn[1], Alexander Eychmüller[2], Karl Leo[1], Sebastian Reineke[1]*

**Affiliations:**
[1] Dresden Integrated Center for Applied Physics and Photonic Materials (IAPP) and Institute for Applied Physics, Technische Universität Dresden; Nöthnitzer Straße 61, 01187 Dresden, Germany

[2] Physical Chemistry, Technische Universität Dresden; Zellescher Weg 19, 01069 Dresden, Germany

[3] Kaunas University of Technology, Department of Polymer Chemistry and Technology; K. Barsausko g. 59, 51423 Kaunas, Lithuania

* Corresponding author. Email: sebastian.reineke@tu-dresden.de

**Abstract:** Wavelength discriminating systems typically consist of heavy benchtop-based instruments, comprising diffractive optics, moving parts, and adjacent detectors. For simple wavelengths measurements, such as lab-on-chip light source calibration or laser wavelength tracking, which do not require polychromatic analysis and cannot handle bulky spectroscopy instruments, light-weight, easy-to-process, and flexible single-pixel devices are attracting increasing attention. Here, we propose a device for wavelength tracking with room-temperature phosphorescence at the heart of its functionality that demonstrates a resolution down to one nanometer and below. It is solution-processed from a single host-guest system comprising organic room-temperature phosphors and colloidal quantum dots. The share of excited triplet states within the photoluminescent layer is dependent on the excitation wavelength and determines the afterglow intensity of the film, which is tracked by a simple photodetector. Finally, an all-organic thin-film wavelength sensor and two applications are demonstrated where our novel measurement concept successfully replaces a full spectrometer.

**Keywords:** Organic room-temperature phosphorescence, colloidal quantum dots, dual-state FRET, organic wavelength sensor, transient photocurrent, off-cycle response.

**Introduction**

Wavelength-sensitive technologies are the basis for both monochromatic and polychromatic spectroscopy and are thus crucial for various fields of application, such as chemical material analysis, characterization of light sources, or calibration of monochromators and laser sources. Established techniques for realizing wavelength-selective light detection use one of two primary approaches: First, wavelength separation by an auxiliary structure (*1*), i.e. a filter array or grating, where narrow light bands are directed onto individual pixels of a photodetector (array) with broadband response. Second, wavelength separation by the photoactive part itself, achieved by a multi-layer arrangement of detectors, where every single detector is sensitive to a specific wavelength band (*2*, *3*). Such spectroscopic devices commonly comprise solid hardware components that restrict them from versatile integration.

Recent developments, however, demonstrate both the huge drive and great potential in terms of easy-to-integrate, low-cost, or lightweight applications (*4*), e.g. narrow-band photodiodes (*5*–*7*), voltage-tunable Fabry-Perot micro interferometers (*8*), or broad-band sensors requiring wavelength multiplexing (*9*). A smart approach to significantly reduce the device complexity was presented by Gautam et al. (*10*). A single-pixel and single-layer device was used to achieve a wavelength-sensitive photocurrent response that can discriminate red, green, and blue (RGB) values via the polarization of a polymer film in an aqueous surrounding. Only recently, a system was presented that consists of a multilayer single pixel of graded-bandgap perovskites evoking a wavelength-sensitive photocurrent response (*11*).

In this manuscript, we present a single-chip wavelength sensor that exploits the dynamics of singlet and triplet states to discriminate a certain input signal. We employ a solution-processed host-guest system comprising organic room-temperature phosphors and fluorescent colloidal quantum dots (QDs), thereby introducing a new, promising application for organic room-temperature phosphorescence (RTP). The latter has been put to multifaceted use (*12*), such as programmable luminescent tags (*13*), oxygen sensing (*14*), or moisture sensing (*15*). Owing to their unique scalable optical properties, high quantum yield, and elevated photostability, colloidal quantum dots proved themselves valuable in a myriad of applications, like light-emitting diodes (LEDs) (*16*, *17*), solar cells (*18*–*20*), or transistors (*21*, *22*). Here, we present a single-layer approach that turns wavelength information into a distinct photocurrent response with a spectral resolution down to 1 nm and below while covering a wavelength range from 300 nm to 410 nm.

## Results

*Materials and conception*

The photoactive layer is drop-cast onto a quartz glass substrate from solution and comprises BP-2TA [4,4'-dithianthrene-1-yl-benzophenone] as RTP emitter (*23*, *24*), CdSe/CdS core-shell QDs as fluorescent emitter (*25*), and a PMMA [poly(methyl methacrylate)] host matrix to suppress non-radiative exciton quenching of the RTP emitter's triplet states (*26*) (**Fig. *1***a). The film thickness ranges around 25 µm (see SI, Materials section for details). The recipe was adopted after screening a range of emitters and concentrations yielding high phosphorescent PL intensity, as presented in Refs. (*24*, *27*).

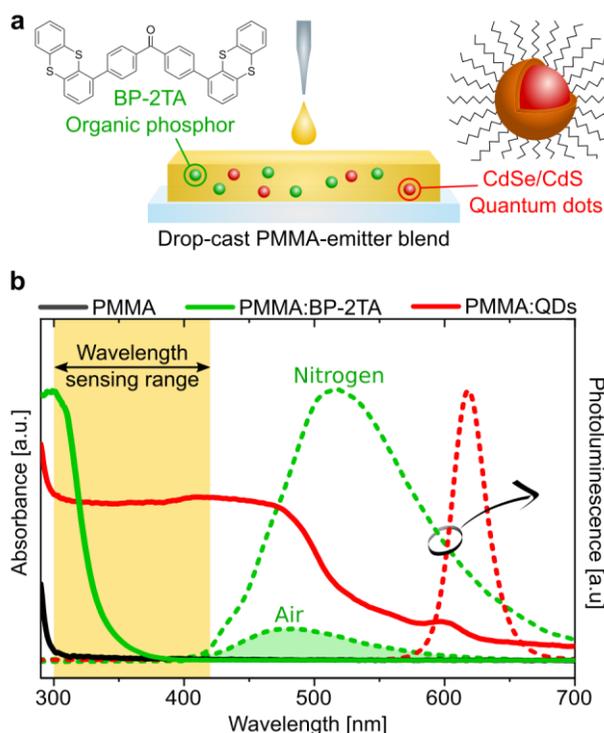

**Fig. 1. Materials and their optical characteristics.** (**a**) Materials used in the wavelength-sensing layer. (**b**) Their respective absorbance (solid lines) and CW photoluminescence emission (dashed lines) profiles. The weak singlet emission of BP-2TA in air (shaded with green) is also shown. Within the indicated range (shaded with yellow), the share of RTP emission under illumination can be tuned by excitation wavelength.

**Fig. *1***b shows the absorbance (solid lines) and photoluminescence (PL) emission profiles (dashed lines) under continuous-wave (CW) excitation at 365 nm of both light-emitting species in PMMA. The host PMMA is not photoactive in the region of interest, as indicated by its flat black absorbance line. The absorbance characteristics of the red-emitting QDs are about constant between 300 nm and 450 nm. Their sharp emission feature just above 600 nm originates from bright excited states that can be understood as being of singlet-like nature, as they show prompt fluorescence (cf. SI, Sect. 1, and Refs. (*25*, *28*)). Despite spin being no appropriate quantum number to describe excited states in semiconductor QDs (*29*), we herein call the origin of the QD emission singlet-like excitons for the sake of discriminating them from the long-living triplet excited states of the green-emitting organic phosphor. The absorbance of the latter peaks at 300 nm and tails off toward 400 nm. Its emission predominantly originates from excited triplet excitons and only a minor contribution can be attributed to its singlet states, as can be seen from its CW PL

emission in air (shaded in green) and nitrogen atmosphere in **Fig. 1**b. According to previous results published from our lab, the fluorescence photoluminescent quantum yield (PLQY) at 365 nm excitation ranges around 1% while its phosphorescence PLQY reaches 20%, which renders the molecule an efficient organic phosphor (*27*).

Absorption of excitation light by BP-2TA produces few singlet ($S_{RTP}$) but mostly triplet excited states ($T_{RTP}$) by efficient intersystem crossing in the steady state. The absorption of excitation light by the quantum dots produces singlet-like excited states ($S_{QD}$). Within the indicated range of wavelength sensing, cf. the yellow shaded area in **Fig. 1**b, the ratio of quantum dot-to-phosphor absorbance is gradually increasing with excitation wavelength. Assuming moderate excitation densities, where nonlinear effects like annihilation processes do not dominate the exciton dynamics, the relative share of the organic triplet exciton density ($T_{RTP}$) to the sum of all luminescent excited state densities ($T_{RTP} + S_{RTP} + S_{QD}$) within the photoactive film, henceforth called the *spin mixing ratio* SMR, becomes thus a function of excitation wavelength:

$$\mathrm{SMR} = \mathrm{SMR}(\lambda_{ex}) = \frac{T_{RTP}}{T_{RTP} + S_{RTP} + S_{QD}}.$$

*Setup and results*

As shown in **Fig. 2**b, the film is excited with a white xenon lamp that is connected to a monochromator, which is set to integer wavelength values. An optical beam shutter cuts off the excitation light to record the off-cycle response of the film. All measurements are conducted at room temperature and under nitrogen atmosphere to prevent triplet quenching by oxygen. The PL intensity of the film is recorded using an amplified silicon photodetector (PD). To exclude any influence of residual UV excitation light, the detector is covered with a 450 nm longpass filter. All data were acquired using Sweep-Me! as control software for automated measurement protocols (*30*).

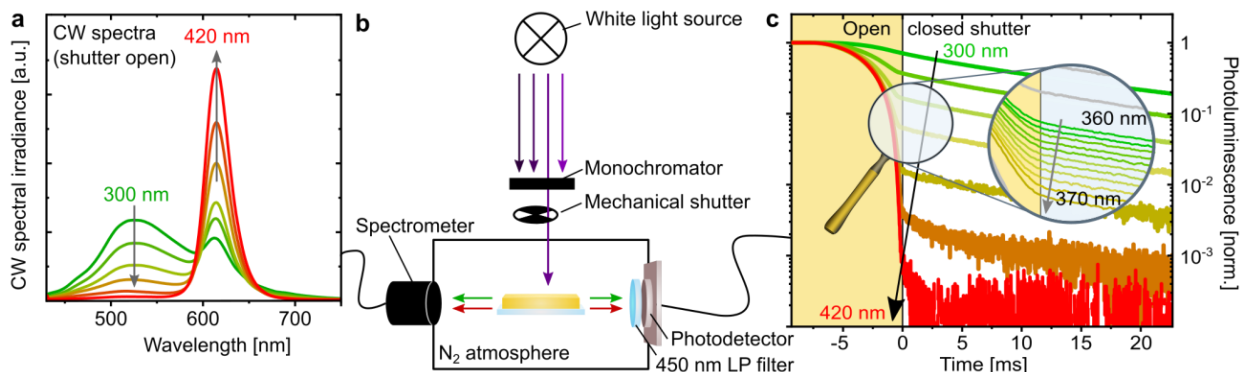

**Fig. 2. Experimental setup for single-film wavelength sensing.** (**a**) PL spectrum of the film depending on the excitation wavelength, trend indicated by arrows. (**b**) Sketch of the experimental setup. (**c**) Specific excitation wavelengths create distinct off-cycle transients. Inset: Even excitation wavelength steps of 1 nm can be discriminated. To keep the figures clear, not all measured transients and spectra are displayed.

The PL of the sample is recorded using a UV-VIS spectrometer for monitoring purposes only and is not required for further data analysis. **Fig. 2**a illustrates how the CW-excited PL spectra depend on excitation wavelength: At 300 nm both the phosphorescent BP-2TA and fluorescent QD emission characteristics are visible (green line), as both molecular species absorb light of this wavelength. With increasing the excitation wavelength toward 420 nm (yellow to red line color) the RTP emission vanishes, as the excitation light only overlaps with the falling flank of its absorbance tail. As a result, the relative share of RTP emission decreases, and the proportion of QD emission increases. Since the triplet decay time of BP-2TA is on the millisecond scale, this luminescence transition can be monitored by detecting the off-cycle PL intensity with a PD. The data is normalized to the average on-cycle PL under CW excitation (shutter open) and thus expresses the share of phosphorescence to total luminescence. **Fig. *2*c** depicts such transients (10 averages each) for an excitation wavelength sweep from 300 nm to 420 nm. The PL plateau is recorded under CW excitation (shutter open). Here, all radiative exciton recombination pathways can be seen ($T_{RTP}$, $S_{RTP}$, and $S_{QD}$). With the shutter being closed, the excitation light ceases and the pure RTP afterglow is visible. Consequently, the onset of the off-cycle PL transient sets a measure for the SMR. With increasing excitation wavelength, the afterglow intensity is reduced, as fewer BP-2TA triplets get excited. At the same time, the QD emission prevails due to their stable absorbance characteristics. As a result, the relative share of excited triplet excitons is reduced and the SMR decreases. Under the condition of moderate excitation intensity and nonlinear exciton dynamics being negligible (see the reasoning in the SI, Sect. 4), exciting the active film with an unknown wavelength within the sensing window thus results in a specific off-cycle transient that can be identified unambiguously with a distinct excitation wavelength.

Two issues exacerbate the readout of the SMR in **Fig. *2*c**: First, the shutter closing takes several milliseconds. Thus, the excitation light quenching does not appear as a sharp edge in the graph and the fluorescence contribution does not vanish immediately. Consequently, the time at which the shutter was fully closed (indicated by the vertical black line in **Fig. *2*c**) was defined as the read-out point for the afterglow intensity. The PL value at this point in time is called *afterglow intensity* and is displayed in **Fig. *3***. Second, the spectral response of the detector needs to be taken into account. The EQE of the silicon PD is not exactly constant over wavelength, i.e. different photon energies are not equally represented in the transient. Because the emission spectrum of the PL film changes with excitation wavelength (cf. **Fig. *2*a**), this evaluation does not produce a true SMR value, but a distorted quantity, earlier defined as *afterglow intensity*. An approach addressing both topics is presented later in this manuscript.

Despite these two issues, a monotonous function for unambiguous wavelength identification is achieved, ranging from 300 nm to about 410 nm (**Fig. *3***). The insets in **Fig. *2*** and **Fig. *3*** indicate that wavelength steps of 1 nm (and even below, see SI, Sect. 6) can be discriminated. Moreover, the system does not suffer from apparent hysteresis, as presented in the SI, Section 2. Figure 3 indicates the upper wavelength-sensing limit (detection limit, gray shaded area), which is set by the PD's specific detectivity, the number of transient averages, and, not least, the properties of the emitter materials blended into the PL film. With our settings chosen as presented, we can discriminate wavelengths up to approximately 410 nm.

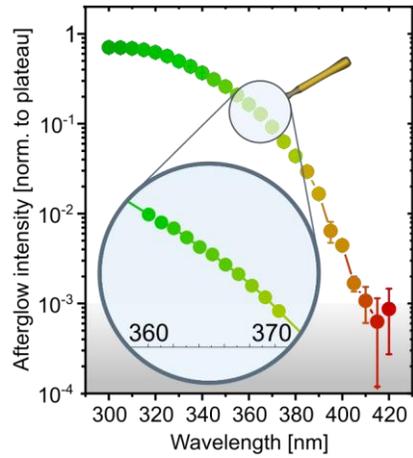

**Fig. 3. Wavelength identification by afterglow intensity.** The afterglow intensity, as defined in the text, is plotted over excitation wavelength and produces a monotonous function. Beyond 410 nm of excitation wavelength, the sensor reaches its detectivity limit (indicated as gray shaded area) and cannot resolve wavelength changes anymore. Error bars indicate the standard deviation of each measurement point.

*Realizing an all-organic wavelength sensor*

Up to now, we have presented a solution-processed single active layer that discriminates wavelengths between 300 nm and 410 nm. Several read-out and device issues have been addressed. We subsequently demonstrate how they can be tackled by changing distinct experimental settings. First, our idea aims at integrated systems, where a thin-film and flexible detector might be favored over a silicon PD. So, we subsequently use a DCV-5T:$C_{60}$ organic photodetector (OPD) instead of the silicon PD, as displayed in **Fig. 4**a. Second, we show that cadmium-based QDs can be replaced by other non-toxic fluorophores, such as organic singlet emitters featuring appropriate absorption profiles. By blending DCJTB [4-(dicyanomethylene)- 2-*tert*-butyl-6-(1,1,7,7-tetramethyljulolidyl-9-enyl)-4*H*-pyran] instead of QDs into PMMA:BP-2TA, the system becomes prone to dual-state Förster resonance energy transfer (FRET): Both the excited triplet and singlet state of BP-2TA can transfer their energy efficiently to the singlet state of DCJTB (*31*). This sensitized fluorescence renders both the film's prompt and delayed emission for all excitation wavelengths to be of the same color (*12*). While the QDs proved unsusceptible to Förster energy transfer from the RTP emitter, DCJTB was seen to accept BP-2TA's triplet and singlet energies quite efficiently (cf. SI, Sect. 1). We suspect the QD's oleic acid ligands together with the host polymer to effectively prevent non-radiative energy transfer by bridging the donor-acceptor distance, which relates to the transfer efficiency by the inverse sixth power (*32, 33*). Hence, DCJTB as singlet-emitting species solves the wavelength-dependent EQE problem of the PD issued above, which would have been of even greater significance when using an OPD.

As the elongated closing time of the mechanical shutter from **Fig. 2**c induces distortions to the PL transients of our film, we now demonstrate that a faster switch-off directly translates into an improved transient signal. Electrically driven LEDs are installed as excitation sources instead. This leads to an immediate drop of excitation light and singlet emission intensity below the millisecond scale, as can be seen in **Fig. 4**b. We obtain fewer observation points on the scale of excitation wavelength since we have only a limited number of LEDs at hand. Now, the plummeting excitation intensity allows to read out a truly proportional SMR value from the onset of the afterglow transient.

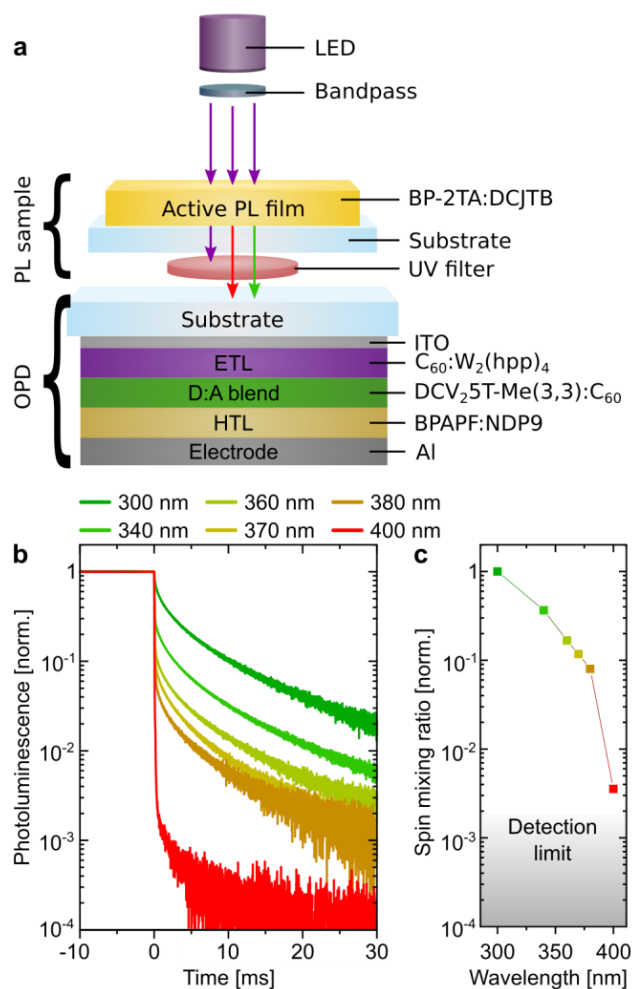

**Fig. 4. All-organic wavelength sensor.** (**a**) Sketch of the sensor stack (not drawn to scale), see Methods section for details. (**b**) The respective normalized transients and (**c**) the detected SMR. A steep singlet emission drop can be observed, owing to the faster switching of electrically-driven LEDs that are used as excitation light sources.

## *Demonstration scenarios for wavelength tracking*

Beyond the conceptual proof of wavelength sensing by exciton spin mixing, two application scenarios are presented in this paragraph, where a simple wavelength tracking system replaces a full spectrometer that would have been necessary otherwise. The experiments use the BP-2TA:QD film connected to the silicon PD equipped with a longpass UV filter.

In the first setup, the 365 nm LED from the former experiment is switched on in continuous mode for heating up. Using a calibrated, commercial spectrometer, we could determine that the LED peak wavelength shifts red by about 0.6 nm over 18 minutes while running at the maximum driving current of 700 mA. The specification sheet of the LED suggests that the case heating due to continuous operation induces this emission shift (see SI, Sect. 8). Performing a calibration measurement, this peak emission shift can be correlated to the PL afterglow intensity of the wavelength sensor (see SI, Sect. 8 for details). A subsequent measurement can now track the LED emission shift using the wavelength sensor only (**Fig. 5**a). After 18 minutes of continuous driving, i.e. heating, the LED is turned off and rests for 30 minutes. Within these 48 minutes, the PL off-cycle response is measured every 2 minutes. A measurement cycle consists of three times switching the LED 300 ms on and 300 ms off and a subsequent averaging of the signals. Every measurement slightly disturbs the pure heating and cooling characteristics of the LED and hence is only repeated three times every 2 minutes. The heating and cooling characteristics of the LED are clearly resolved by our sensor.

For the second demonstration, a pulsed and tunable dye laser is used as excitation source. Again, a calibration measurement enables an experiment, where the introduced wavelength sensor can be used to determine the emission wavelength of the laser (**Fig. 5**b). The laser was tuned back and forth while measuring the PL response of the active film 10 times and averaging. As the laser pulse is only about 1 ns long, the triplet density has only little time to build up. Hence, the afterglow intensity is very weak. this yields a high standard deviation limiting the resolution to well above 1 nm at 10 averages, although the laser wavelength is in the sweet spot of the system's resolution and the excitation intensity is quite high. The short excitation pulse does not produce a stable PL plateau to which the afterglow data can be normalized. Instead, it is normalized to the integral of the detected prompt PL burst of the film. Hence, the units of the afterglow intensity are only arbitrary. Details can be found in the SI, Sect. 9.

Both proof-of-concept experiments show clearly that the exciton spin mixing system is well capable to track the spectral intensity of a given light source and suggest that it can be particularly useful for simple monitoring tasks. Here, one will benefit from the low system complexity and cost.

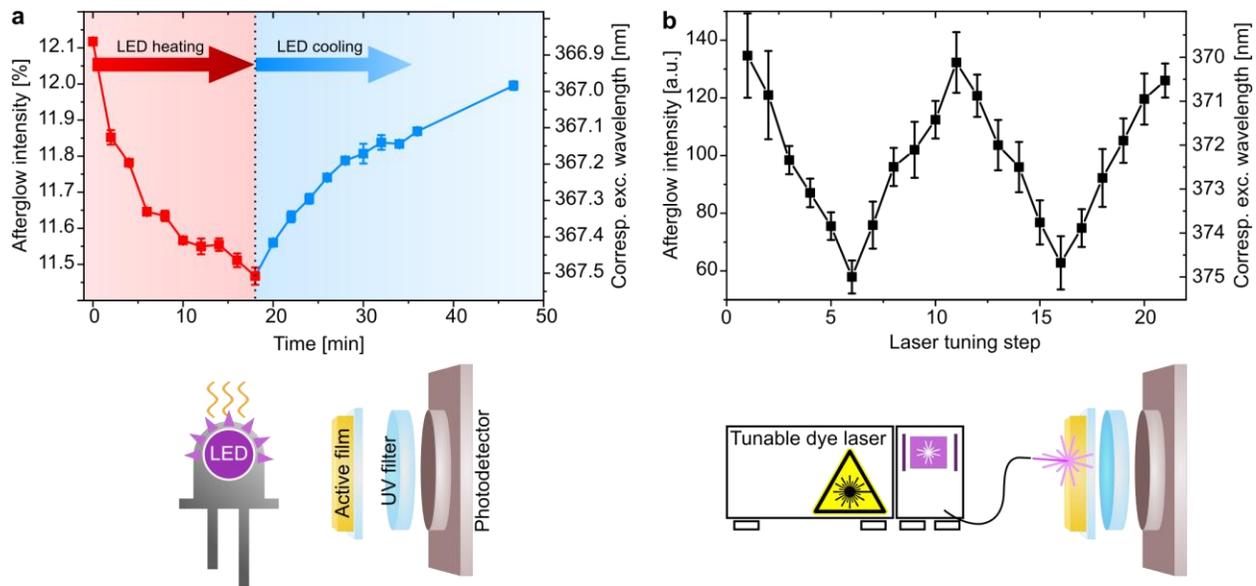

**Fig. 5. Demonstration of wavelength tracking applications.** (**a**) Tracking of LED heating as well as cooling and the corresponding wavelength shift. (**b**) Wavelength tracking of a tunable dye laser. The error bars indicate the standard deviation of the values. Details can be found in the SI, Sect. 8 and 9.

**Discussion and Perspectives**

The two demonstrated tracking applications require calibration with the very same excitation source. First, the spectral shape of the light source influences the afterglow intensity characteristics, as not only the peak wavelength may determine the off-cycle response. As we present a simplistic one-dimensional readout scheme, our sensor cannot discriminate between different spectral shapes. Second, the exposure time of the excitation source is crucial for whether the PL film reaches a steady state of excited triplet excitons. Any exposure time well above the lifetime of the triplet excitons will not change the afterglow intensity. However, as presented in the laser tracking experiment, very short exposure times drastically reduce the afterglow intensity and hence the signal-to-noise ratio. In case of monochromatic excitation and long exposure times, e.g. with a CW laser and shutter combination, where those issues do not arise, the calibration source can be different from the test source.

For the presented measurement concept, the boundaries of the sensitive interval and the spectral resolution are set by the absorbance characteristics of the fluorophores and phosphors embedded into the film. With the presented materials, we can scan an interval ranging from 300 nm to 410 nm and achieve a monotonously falling afterglow intensity, which in turn enables discrimination of excitation wavelengths within that range. In the steepest range of the afterglow intensity profile, a resolution down to one nanometer or below is achieved, depending on the experimental settings, like excitation intensity and averages (SI, Sect. 6). Using phosphorescent molecules with red-edge absorption tails in different wavelength regimes, such as $BF_2(HPhN)$ (*34*), $BF_2$bdks (*35*), or PhenTpa (*36*), adopts the sensitivity range of the system according to the experimental purpose. Potential application scenarios, such as monochromator calibration or laser wavelength tracking, might require high spectral resolution. This quantity relates closely to the absorption profile gradient of the RTP emitter (SI, Sect. 6). For such scenarios, it is feasible to employ phosphors with steep absorption edges, e.g. Spiro-TAD (*37*), Spiro-TTB (*38*), or Spiro-OMe-TAD (*39*), which are established, temperature-stable OLED materials (*40–42*). Also, the optimized emitter concentration can change the required exciton dynamics, as scanned in Ref. (*24*). Sensitivity range and resolution are hence trade-offs that can be adjusted according to the particular application concept.

Fluorescent quantum dots are easily scalable throughout the visible wavelength range and are therefore adaptable to any phosphorescent molecule, making them our composite of choice. To demonstrate an all-organic and cadmium-free device for versatile integration, we used a sensor employing an organic PD and organic fluorophore. It is also worth stressing that the wavelength-sensitive film itself can be physically decoupled from the electrically wired (O)PD. For applications in challenging surroundings, this might pay off. By chemical passivation or encapsulation, molecular emitters were also demonstrated stable under moist or ambient conditions (*43–46*).

As mentioned above, the definition of the SMR is only valid while non-linear effects are negligible. Using a high power LED, we scanned a wide range of excitation intensity and determined the excitation sources used throughout the manuscript to cause no significant non-linearity (SI, Sect. 4). This finding is accounted for by three main points: A heavily diluted RTP emitter (3 wt%), the polymer matrix, and finally the low triplet lifetime (30 ms), all of which causing a low triplet-triplet interaction probability. For further investigations comprising different RTP emitters, however, the onset of annihilation-induced nonlinearities above a certain excitation intensity threshold needs to be considered.

The most intense transients take about 250 ms to reach the detector's noise level with the material system used in our experiments. The PL film requires about the same time to achieve a steady level of excited triplet states under illumination. Hence, a rectangular light source driving voltage could be run at approximately 2 Hz. This parameter, however, strongly depends on the phosphor's triplet lifetime and the readout procedure. Using an RTP emitter with a shorter lifetime in the microsecond range, e.g. PtOEP, would significantly enhance the potential readout speed. For data acquisition, we averaged every transient 10 times. Another way could be the integration of a single transient. This reduces the susceptibility to noise, as not only one single value per transient is measured. This approach could be beneficial in terms of readout speed since less or no averages need to be taken (cf. SI, Sect. 3). In contrast to the readout procedure in the time domain presented in this text, a dual-wavelength OPD (*47*) could be used to identify the PL contributions of singlet and triplet emitters, and hence determine the spin mixing ratio.

The spectral detectivity of the complete sensor system relates to parameters like the PLQY of the emitting molecules, incoupling efficiency of the film's PL into the PD, and finally the specific detectivity of the PD itself. It, therefore, offers a number of skrews to be turned in order to optimize the system for specified use cases. The presented wavelength tracking routine applies to a scenario where the excitation can be ceased on the millisecond scale using a shutter or an electric switch.

The wavelength sensor does not require any additional refractive elements. We had, however, to apply a wavelength longpass filter in front of the PD to cut off the remaining excitation light (cf. SI, Fig. S2b). For a monolithic solution, we are currently developing thin-film filters that can be integrated into our thin-film processing techniques. Otherwise, either a suitable substrate may act as an excitation filter, or a grazing-angle experiment could be realized that limits the amount of excitation light. Moreover, we already presented efficient oxygen barrier layers that enable RTP applications in ambient atmosphere and can also be applied here (*24*).

**Conclusion**

This article presents the ability of wavelength sensing using a single PL film attached to a simple broad-band detector. The pivotal idea is to specifically exploit the overlapping absorption profiles of fast- and slow-emitting molecules that are blended into the PL film, i.e. to selectively excite rather singlet or triplet excited states and, therefore, to control the spin mixing of the emitter system. By tracking their PL response, the readout is therefore solely conducted in the time domain and does not require diffractive elements. The presented sensor can scan a spectral range of about 100 nm with a maximum resolution below one nanometer. It does not compete with full spectrometers but can facilitate wavelength measurements in integrated systems, such as monochromator calibration or laser wavelength tracking. An all-organic solution based on an organic PD is presented, demonstrating how such a sensor can be manufactured entirely from versatile components. We further discuss how the choice of emitting materials sets the frame for adapting spectral resolution and sensitive range for individual purposes and finally demonstrate heating-induced LED emission shifts and laser wavelength tracking as two application scenarios.

**Acknowledgments:** The authors thank Axel Fischer for contributing Sweep-Me! device classes that were necessary for automated measurement protocols.

**Funding:** This work received funding from the European Research Council under the European Union's Horizon 2020 research and innovation program (Grant Agreement No. 679213; project acronym BILUM). A.K. received funding from the Cusanuswerk Foundation and acknowledges funding from the DFG project HEFOS (Grant No. FI 2449/1-1). J.B. acknowledges the DFG project VA 1035/5-1 (Photogen) and the Sächsische Aufbaubank through project no. 100325708 (InfraKart).

**Author contributions:**

    Conceptualization: AK, SR

    Methodology: AK, TB, TA, JB, FF

    Investigation: AK, TB, RW, FF, TA, MG, CG, JB, AT

    Visualization: AK, MG

    Funding acquisition: AK, JB, KL, AE, SR, MG

    Project administration: AK, SR

    Supervision: AE, KL, SR

    Writing – original draft: AK, JB

    Writing – review & editing: all

**Conflict of interests:** The authors declare no competing interests.

**Data and materials availability:** All data needed to evaluate the conclusions in the paper are present in the paper and/or the Supplementary Materials. Additional data related to this paper may be requested from the authors upon reasonable request.

**Supplementary Materials**
Materials and Methods
Supplementary Text
Fig. S1. Lifetime measurements of the investigated blends.
Fig. S2. Photoluminescence spectra of the relevant material systems.
Fig. S3-1. Statistics and hysteresis of wl. tracking performance of the BP-2TA:QD system.
Fig. S3-2. Statistics and hysteresis of wl. tracking performance of the BP-2TA:DCJTB system.
Fig. S4. Afterglow intensity measurement by integration.
Fig. S5. Linear dynamic range of the investigated PL film.
Fig. S6. Relation between absorbance profile and sensing resolution.
Fig. S7. Temporal intensity characteristics of the excitation light sources.
Fig. S8. Intensity profile of the white light source and shutter combination.
Fig. S9. Wavelength tracking of LED heating.
Fig. S10. Laser wavelength tracking.
Tab. S1. Optical intensities of excitation sources.
Tab. S2. Device structure of the OPD.

# Supplementary Materials for

# Accurate wavelength tracking by exciton spin mixing

Anton Kirch, Toni Bärschneider, Tim Achenbach, Felix Fries, Max Gmelch, Robert Werberger, Chris Guhrenz, Aušra Tomkevičienė, Johannes Benduhn, Alexander Eychmüller, Karl Leo, Sebastian Reineke*

Correspondence to: sebastian.reineke@tu-dresden.de

**Table of contents**

**List of figures and tables**

## 0. Materials and Methods

Film preparation

The active film consists of a simple host-guest system that was drop-cast from solution onto an inch-by-inch quartz glass substrate. We used PMMA [poly(methyl methacrylate)] as host material (average molecular weight 550000 u, purchased from Alfa Aesar) dissolved in anisole (80 mg/ml). The RTP emitter BP-2TA [4,4'-dithianthrene-1-yl-benzophenone] was synthesized as described in Ref. (*1*) and dissolved in anisole (10 mg/ml). The organic fluorophore DCJTB [4-(dicyanomethylene)- 2-tert-butyl-6-(1,1,7,7-tetramethyljulolidyl-9-enyl)-4H-pyran] (purchased from Sigma Aldrich) was dissolved in anisole (10 mg/ml). The CdSe/CdS quantum dots were synthesized as described in Ref. (*2*) and dissolved in toluene. All solutions were stirred for 2 hours on a hotplate set to 50°C. For preparing the final solution, 400 µl of PMMA solution (58 nmol of PMMA, > 96 wt%), 100 µl of BP-2TA solution (1625 nmol of BP-2TA, 3 wt%) and either 5 µl of QD solution (125 pmol of QDs) or 10 µl of DCJTB solution (220 nmol of DCJTB, 0.3 wt%) were mixed. After further stirring, the films were drop-cast and left for drying in a fume hood overnight.

Spectroscopic and photocurrent data acquisition

All data were acquired under nitrogen atmosphere**.** The edge emission spectra and photo transients were taken using a UV/VIS spectrometer (CAS 140CTS, Instrument Systems) and an amplified silicon photodiode (PDA100A, Thorlabs) or self-fabricated organic photodiode with a current amplifier (Femto, DLPCA-200) using a multifunctional lab instrument (STEMlab 125-14, Red Pitaya) and Sweep-Me! as control software for automated measurement protocols (*3*). The excitation source was operated for 300 ms (on-cycle) and rested for another 300 ms (off-cycle) for every measurement. This is 10 times the triplet lifetime of the organic phosphor and ensures a stable triplet density. The PL transients were averaged 10 times. As a light source, a xenon white-light source (LOT Quantum Design LSH-302) and monochromator (Bentham MSH-300, Quantum Design) with an attached optical beam shutter (SH05, Thorlabs) were used. The monochromator uses two gratings within the wavelength region of interest ($\lambda \leq 395$ nm: MSG-T-2400-250 with 2400 l/mm, $\lambda > 395$ nm: MSG-T-1200-500 with 1200 l/mm). The specifications for the 1200 l/mm grating are resolution < 0.3 nm, accuracy ± 0.2 nm, and reproducibility ± 0.05 nm (*4*). In the case of the all-organic sensor, four different mounted LEDs (Thorlabs **M300L4,** M340L4, M365L2, M405L3) were used. Except for the 300 nm LED, all of them were equipped with respective bandpass filters (FB340-10, FB360-10, FB370-10, FB380-10, FB400-10 from Thorlabs) to achieve six excitation wavelengths. Absorbance spectra were taken using a spectral photometer UV-3100 (Shimadzu Deutschland). The 337 nm nitrogen laser (MNL 200, 200 kW excitation pulse power, 1 Hz frequency, 700 - 1000 ps pulse length) was fabricated by Lasertechnik Berlin and equipped with a tunable dye module (ATM, Butyl-PBD in toluene, concentration of $4\cdot10^{-3}$ molar).

OPD fabrication

The layers of the OPDs are thermally evaporated in a vacuum system (Kurt J. Lesker Company Ltd., USA) at ultrahigh vacuum (base pressure of $< 10^{-7}$ mbar) on a glass substrate with a pre-structured indium-tin-oxide (ITO) contact (Thin Film Devices). The substrate is cleaned in a multistep wet process including rinsing with N-methyl-2-pyrrolidone, ethanol, and deionized water as well as treatment with ultraviolet ozone. Details on the layer sequence are listed in Table S1, Sect. 5, SI. All organic materials were purified one to two times by gradual sublimation. The device area is defined by the geometrical overlap of the bottom and the top contact and equals 6.44 mm$^2$. To avoid exposure to ambient conditions, the organic part of the OPD is covered by a small, transparent glass substrate, which is glued on top by employing a UV light curing epoxy resin (Denatite XNR 5592, Nagase & Co. Ltd, Japan).

## 1. Investigation of Förster resonance energy transfer (FRET)

Fig. S1 presents data demonstrating the BP-2TA:QD system to be less susceptible to FRET than the BP-2TA:DCJTB system. Both systems are evaluated via time-correlated-single-photon-counting (TCSPC, PicoQuant TimeHarp, 367 nm excitation) at 475 nm (singlet emission of BP-2TA) under acceptor concentration variation. According to Förster's theory of energy transfer, an increase in acceptor concentration yields a higher Förster transfer rate and thus a more rapid depletion of the donor's singlet excited state. As can be extracted from Fig. S1a and S1b, DCJTB is acting as acceptor within our settings while the QDs are not. The same holds for the triplet transients, investigated with a silicon PD in the same experimental scenario as described in the main manuscript. We, therefore, deduce that both efficient singlet-singlet and triplet-singlet FRET exclusively appear in the BP-2TA:DCJTB blend. Apparently, the combination of PMMA as molecular matrix and oleic acid ligands effectively screens the QDs from non-radiative energy transfer.

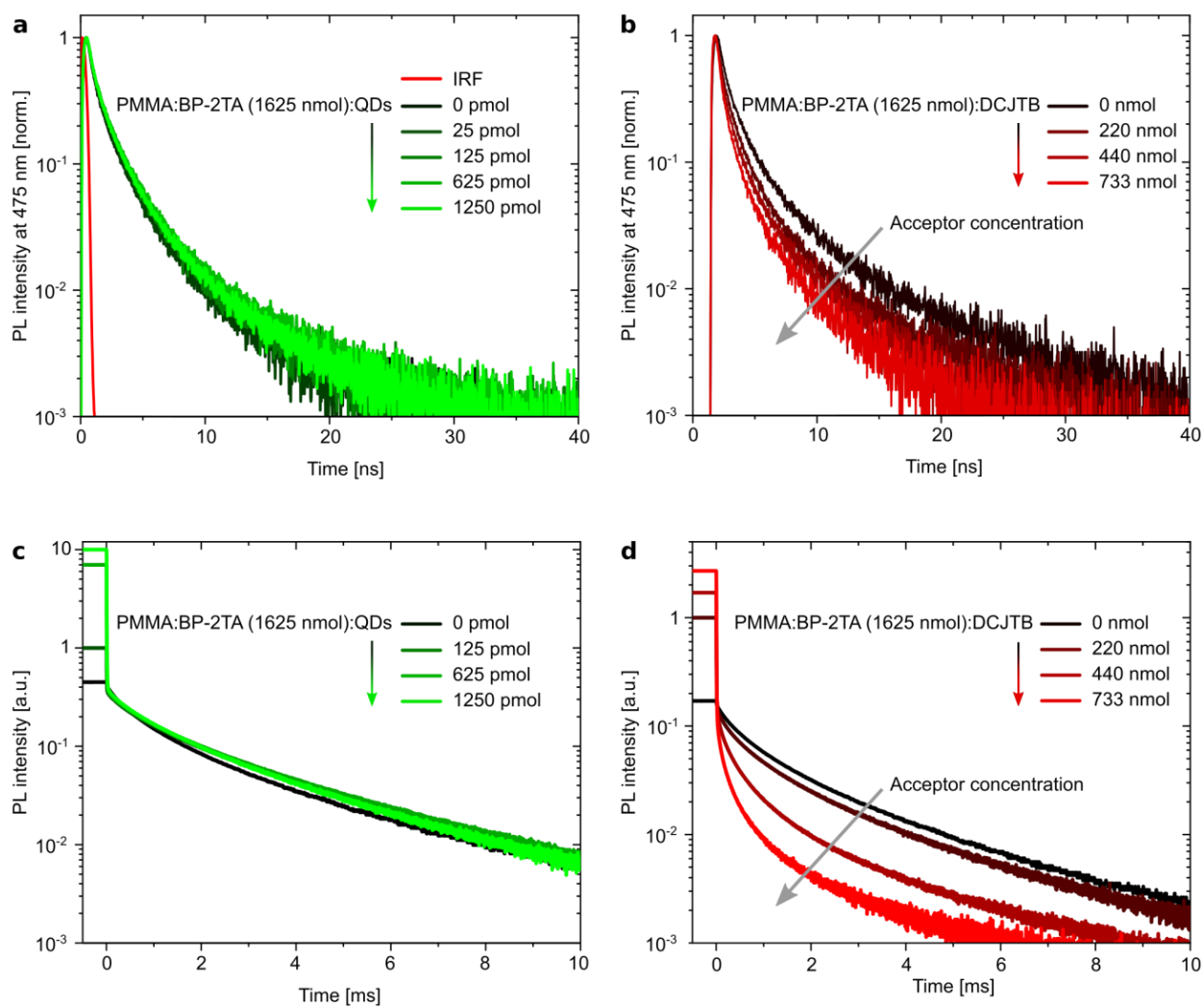

*Fig. S1. Lifetime measurements of the investigated blends. IRF is the Instrument Response Function, which is the excitation laser pulse in this case. (a) and (b) show TCSPC measurements to evaluate BP-2TA's singlet lifetime for the QD and DCJTB blend, respectively. (c) and (d) show BP-2TA's triplet transients recorded with the silicon photodiode PDA100.*

Fig. S2a and S2b show CW spectra, Figures S2c and S2d delayed spectra of the BP-2TA:QD system (left) and BP-2TA:DCJTB system (right). As becomes apparent from Fig. S2a, the CW emission color of the blend comprising QDs shifts with excitation wavelength. Conversely, the DCJTB blend emission spectrum appears unsusceptible to the excitation wavelength, as a significant share of both the singlet and triplet excited state energies of BP-2TA are transferred to the singlet states of DCJTB, which subsequently recombine radiatively. The weak QD emission in the afterglow spectra, and therefore also part of the QD emission in the CW spectra, we assign to radiative energy transfer (reabsorption).

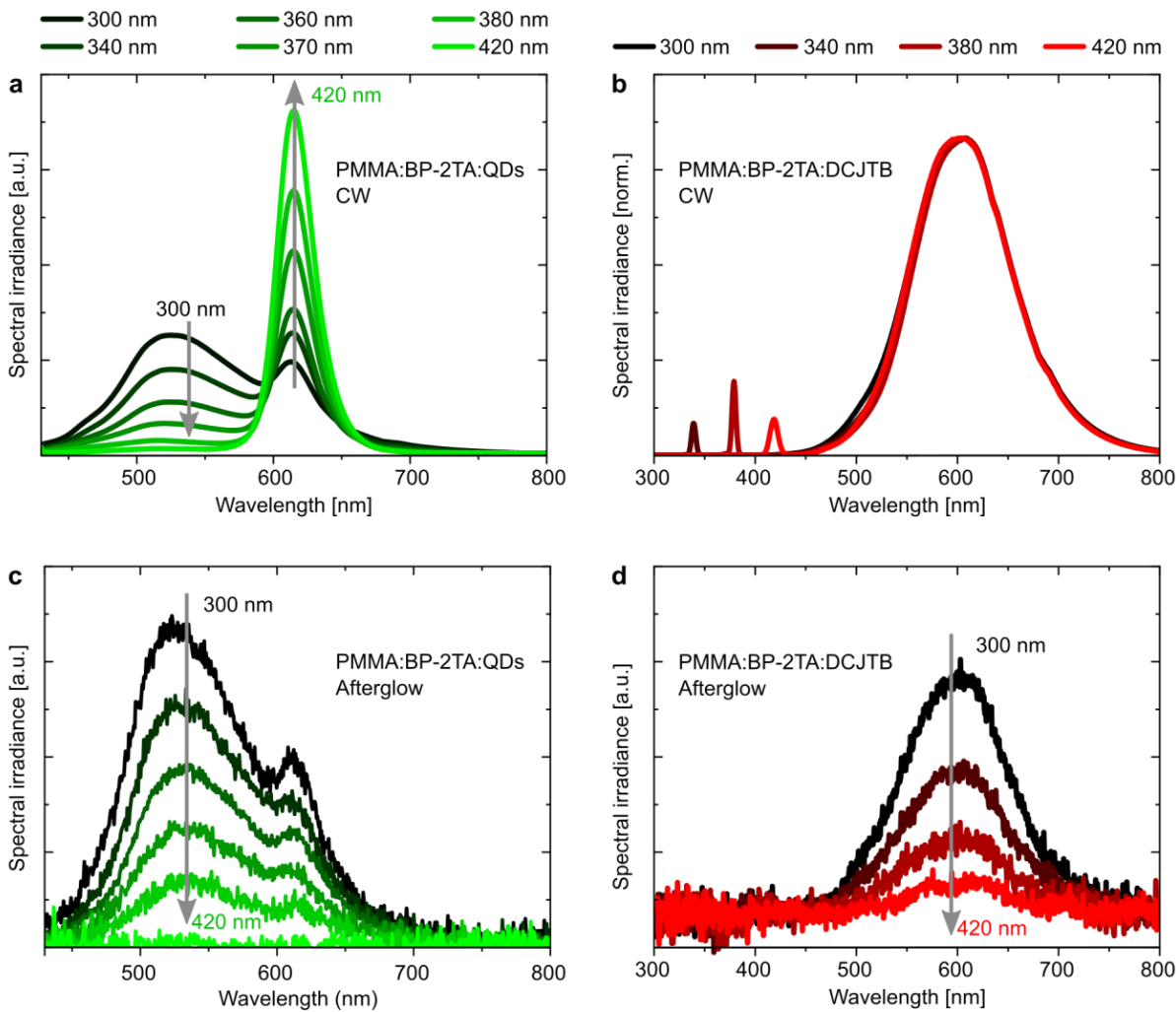

*Fig. S2. Photoluminescence spectra of the relevant material systems. (a) and (b) show CW spectra, (c) and (d) afterglow spectra of the blended systems used throughout the experiments in dependence of excitation wavelength. While both emission contributions can be seen from the QD system, DCJTB acts as efficient acceptor to both BP-2TA's singlet and triplet excitons, rendering the blend's emission color purely red.*

## 2. Hysteresis measurements

Hysteresis can be introduced by non-uniform photodegradation of the emitting sites in the blend, i.e. if the RTP molecule degrades faster than the colloidal QDs. With the excitation energies used in our experiments, we could not observe such a phenomenon on a timescale relevant to our measurements (cf. Figures S3a and S3b). The roundtrip scan took about 2 hours. We could, however, identify a slight drift in the response of the shutter to the trigger signal (< 1ms/hour), which might be caused by the heating of mechanical components. The measurement was triggered by the electrical signal which drives the shutter. We corrected the transients in time to the falling flank of the excitation light.

Figures S3A and S3B present an elaborate study on the hysteresis and error margins of the sensor systems used in the main manuscript. Figure S3a shows the BP-2TA:QD system scanned with the white light source + monochromator system as excitation source, while Figure S3b presents the same measurements taken with the BP-2TA:DCJTB system excited with LEDs.

Figure S3c plots the relative deviation $d_x$ between forwarding $x_f$ and backward $x_b$ measured points:

$$d_x = 2\frac{x_{\text{backward}} - x_{\text{forward}}}{x_{\text{backward}} + x_{\text{forward}}}.$$

Beyond 390 nm of excitation wavelength, where the afterglow signal becomes very low, the relative deviation increases drastically. The same information is carried by Figure S3d, which plots the relative standard deviation of the forward wavelength sweep. As stated in the main text, 10 transient averages were taken at each wavelength step.

The weaker the PL of the RTP molecule, the higher the relative standard deviation of each measurement point. The main difference between both excitation systems, apart from the step widths of the excitation wavelength, is the intensity of the excitation light. As summarized in Tab. S1, the LEDs provide much higher excitation intensities and can hence significantly improve the signal-to-noise ratio and sensitivity of the detector system. In this way, even wavelengths beyond 400 nm can be discriminated and both the relative deviation and standard deviation at long wavelength are reduced. A second effect in the mid-range of the sensitive spectrum is visible: While the relative deviation up to 390 nm is low for the first presented scenario, we see a significant error for the LED-excited system already at 340 nm – 370 nm. We do not account this, however, to a high fluctuation in measurement data (the standard deviation in Fig. S3-2a and b are barely visible) but to the sensibility of the system to the excitation sources themselves. As discussed in the main manuscript and SI, section 8, the measurement is highly sensitive to the thermal conditions of the LEDs, especially in the range of the rapidly dropping absorption characteristics of BP-2TA between 330 nm and 370 nm. And this is what most probably overlais the otherwise monotonically increasing relative standard deviation in Figs. S3-2c and d.

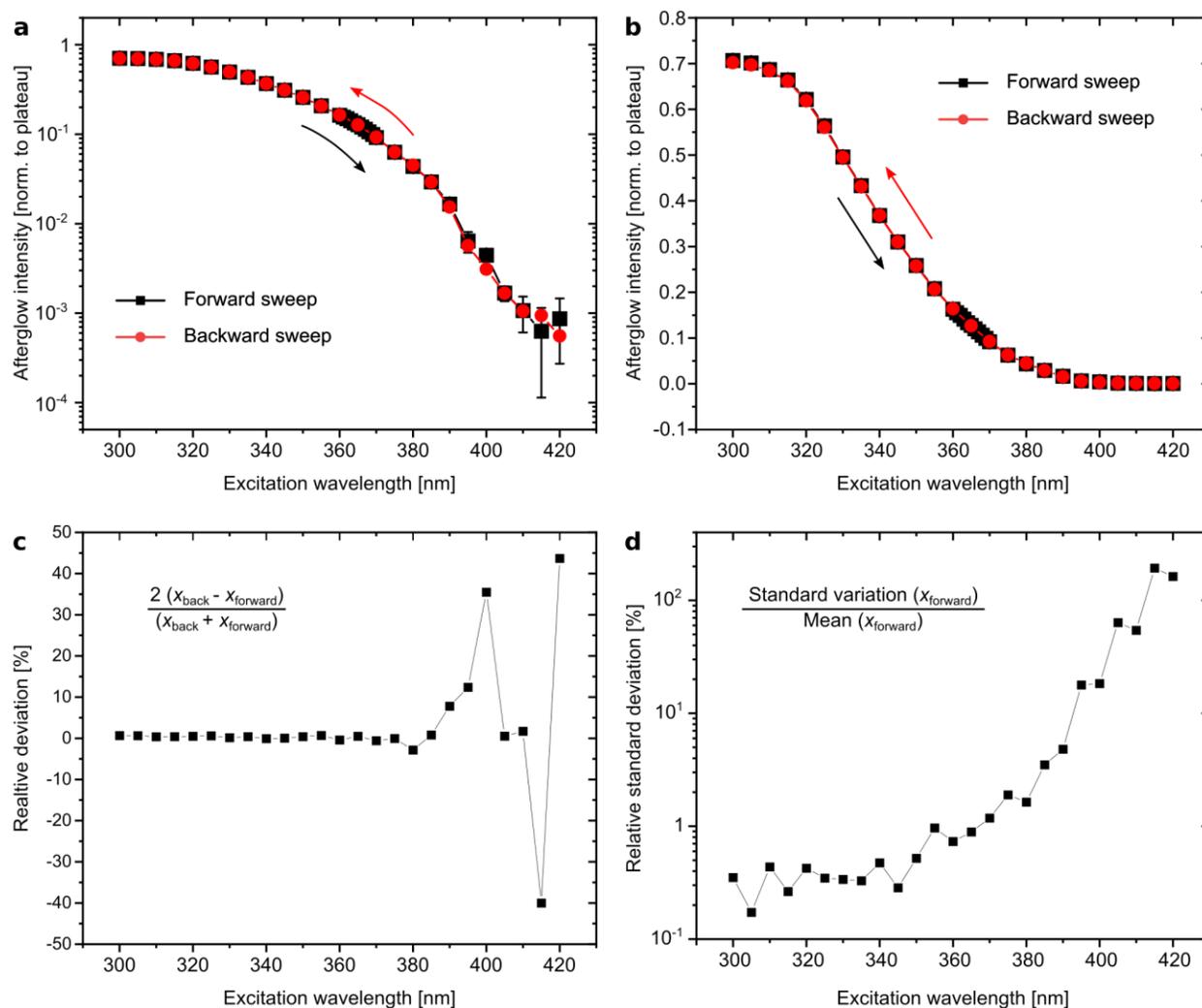

*Fig. S3-1. Statistics and hysteresis of wavelength tracking performance of the BP-2TA:QD material system. Hysteresis of the system's afterglow intensity in (a) logarithmic and (b) linear scale taken with the white light source + monochromator combination. Error bars indicate the standard deviation for each set of measurements (10 averages). (c) Shows the relative deviation of forward and backward sweep. (d) Gives the relative standard deviation of the forward sweep, which scales with the reduced afterglow intensity.*

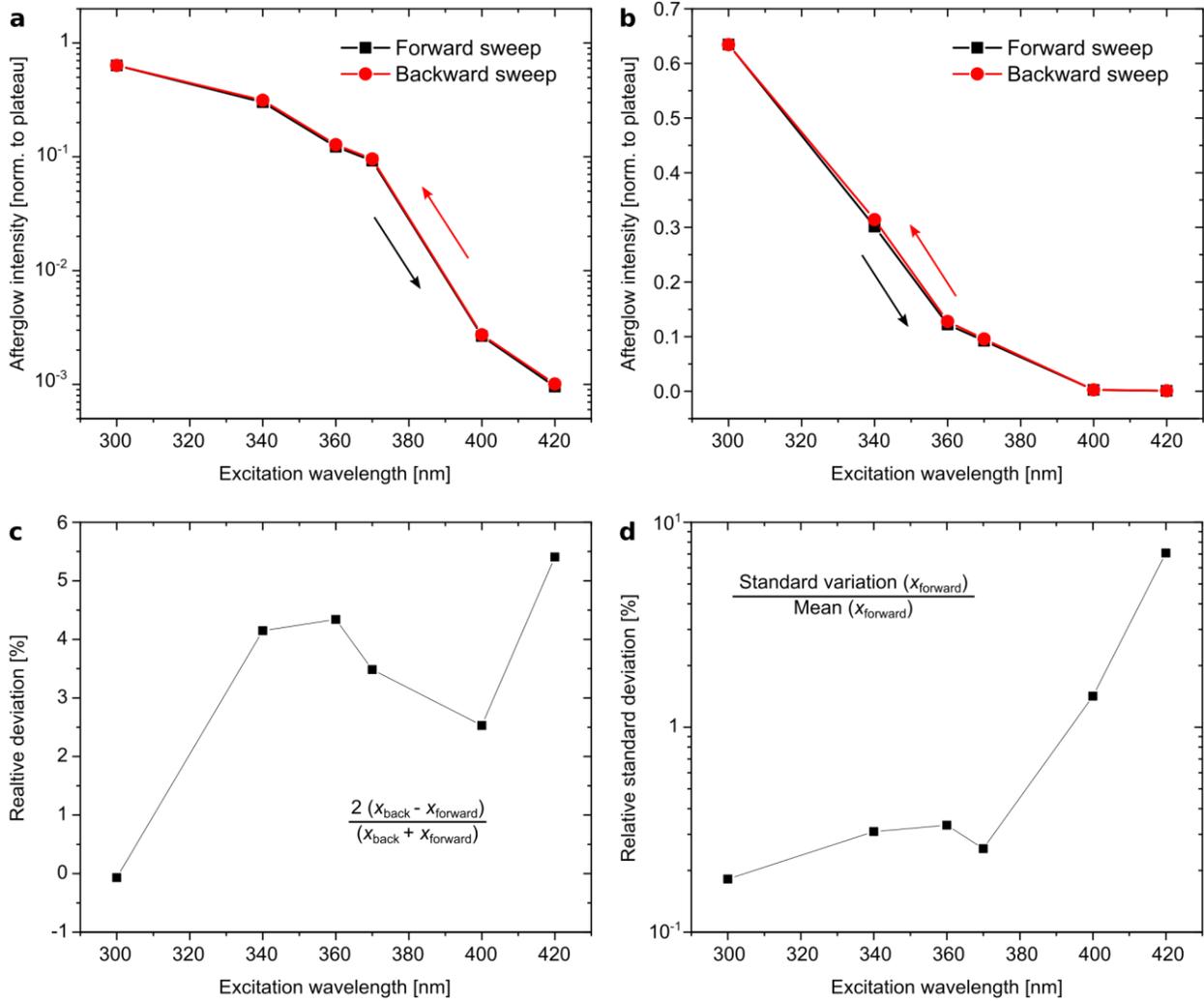

*Fig. S4-2. Statistics and hysteresis of wavelength tracking performance of the BP-2TA:DCJTB material system. Hysteresis of the system's afterglow intensity in (a) logarithmic and (b) linear scale taken with LEDs as excitation sources (10 averages). Error bars indicate the standard deviation for each set of measurements. Plot (c) shows the relative deviation of forward and backward sweep. Plot (d) gives the relative standard deviation of the forward sweep, which scales with the reduced afterglow intensity.*

### 3. An alternative approach to obtain the afterglow intensity

Throughout the manuscript, we read out the afterglow intensity by a single value (readout point = onset of the afterglow transient when the excitation light is completely off). This routine is beneficial in terms of computing capacity and effort, as only one single value needs to be considered. Another possibility, however, could be the numerical integration of the entire transient signal following the readout point (cf. Figure S4a). That gives a measure being less susceptible to signal fluctuations and noise and hence could reduce the number of transient averages required to get clear information. Just like the former algorithm, it yields a monotonous function of afterglow intensity over excitation wavelength (cf. Figure S4b). As the afterglow intensity at 420 nm is close to zero, a randomly negative integral was calculated here, which is why the value is not displayed in the logarithmic plot (missing data point in Fig. S4b at 420 nm).

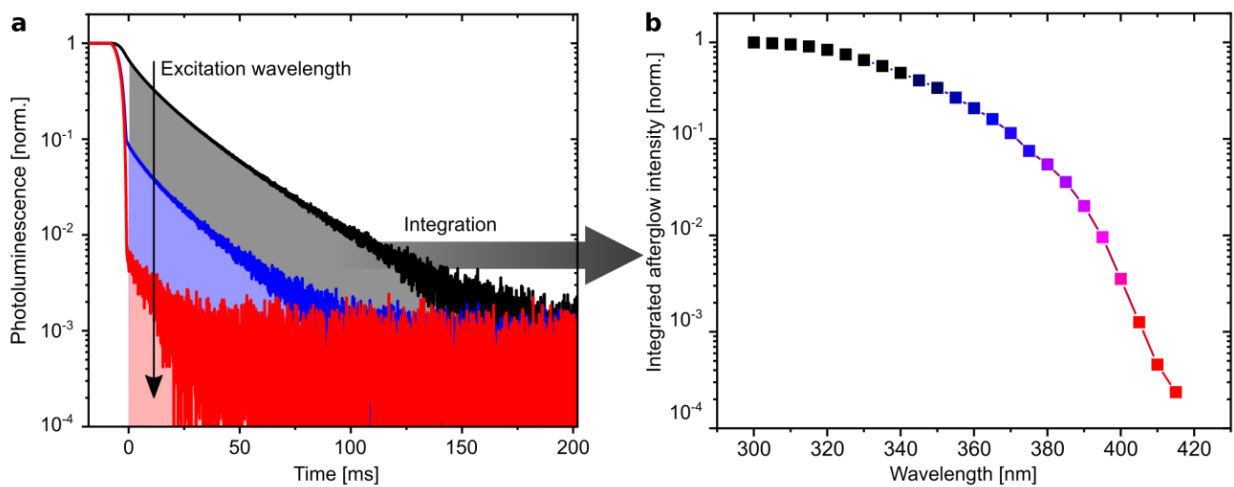

*Fig. S4. Afterglow intensity measurement by integration. (a) Depiction of transient integration yielding (b) an integrated afterglow intensity that is again a monotonous function of excitation wavelength.*

## 4. Linear dynamic range and excitation intensity scan

Changes in the excitation intensity can alter the exciton dynamics in PL films. Increasing the excitation intensity may lead to more annihilation processes, as the density of excited states increases. To prevent annihilation processes, we diluted the RTP emitter at a very low concentration (3 wt%) into the host matrix (> 96 wt%). Both the low emitter concentration and the polymer matrix prevent quenching. Additionally, the triplet lifetime of BP-2TA is rather short, which again reduces the probability of triplet-triplet interaction.

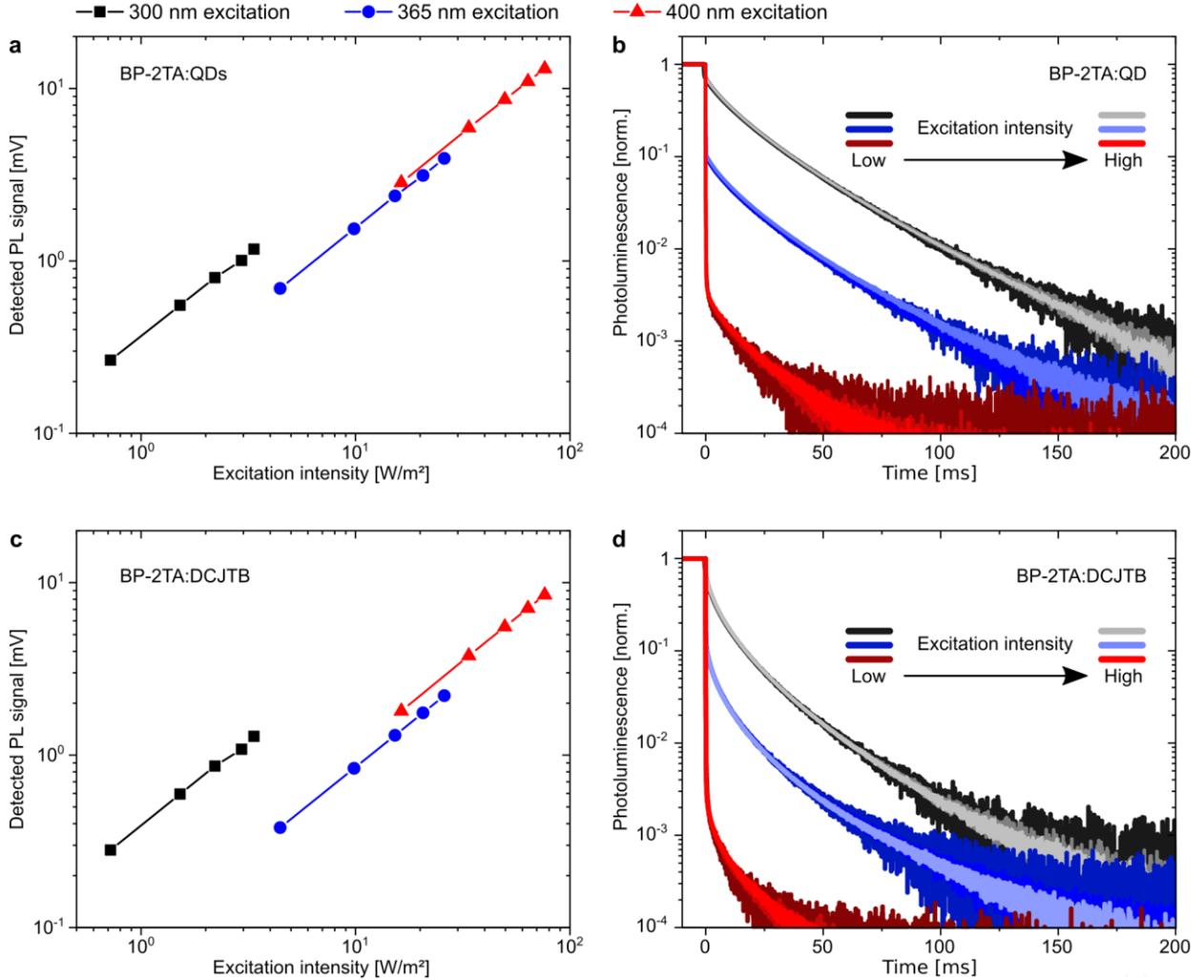

*Fig. S5. Linear dynamic range of the investigated PL film at three different excitation wavelengths. (a+c) The detected on-cycle PL intensity of the respective film depending on the excitation intensity. (b+d) The afterglow transients measured for the three excitation wavelengths and five excitation intensities given in a and c (dark to bright line color indicates increasing excitation intensity). (a+b) Show data for the BP-2TA:QD materials system and (c+d) for the BP-2TA:DCJTB material system.*

To investigate whether the excitation densities applied throughout our experiments cause significant annihilation processes, we used three of the LEDs also used in the main manuscript. The excitation intensity illuminating the PL sample depending on the LED driving voltage (1V to the maximum of 5V) was scanned by mounting a spectrometer head (calibrated spectrometer CAS 140CTS, Instrument Systems) in place of the sample. We subsequently measured the PL transient of the film with the silicon PD. We tested this behavior for both the BP-2TA:QD and the BP-2TA:DCJTB system. The detected on-cycle PL signal scales linearly with excitation intensity (cf. Figure S5a and S5c). Moreover, we measured the off-cycle transients of the PL film for the same range of excitation intensities and could not detect any change in their shape (cf. Figure S5b and S5d). If annihilation processes were significantly influencing the exciton dynamics in the film, one would expect a nonlinear excitation-PL intensity relation and an intensity-dependent triplet lifetime. As both scenarios are not observed for either blend, we conclude that our excitation sources do not cause severe annihilation processes for our material systems.

We used the 300 nm LED without bandpass, the 340 nm LED with 340 nm bandpass filter, 365 nm LED with 360 nm and 370 nm bandpass filters, and the 405 nm LED with 400 nm bandpass filter as specified in the Methods section.

Both the LED-filter combinations and the white-light source combined with the monochromator feature excitation intensities that change with wavelength. Table S1 lists their intensities at specific excitation wavelengths.

*Table S1. Optical intensities of excitation sources. The optical power densities measured with a calibrated UV/VIS spectrometer at the position of the PL film in the setup for different excitation sources. We used the white xenon source with a monochromator at different wavelengths and several LEDs with bandpass filters attached.*

| Light source | Wavelength [nm] | Optical intensity at sample position [W/m²] |
|---|---|---|
| White xenon source + monochromator | 300 | 0.002 |
| | 320 | 0.022 |
| | 340 | 0.380 |
| | 360 | 1.239 |
| | 380 | 1.176 |
| | 400 | 5.047 |
| | 420 | 6.512 |
| LED + band-pass filter | 300 | 3.4 |
| | 340 | 6.6 |
| | 370 | 25.9 |
| | 400 | 77.7 |

## 5. Organic photodiode stack

*Table S2. Device structure of the OPD.*

| Material (chemical name) | Function | Thickness, Processing condition | Supplier |
|---|---|---|---|
| Al | Anode | 100 nm | Kurt J. Lesker Ltd., USA |
| NDP9 | Injection layer | 1 nm | Novaled AG, Germany |
| BPAPF:NDP9 | Hole transporting layer | 40 nm, 10 wt% NDP9 | *See other rows* |
| BPAPF (9,9-bis[4-(N,N-bis-biphenyl-4-yl-amino)phenyl]-9H-fluorene) | Electron blocking layer | 5 nm | Lumtec, Taiwan |
| $DCV_2$5T-Me(3,3):$C_{60}$ ($DCV_2$5T-Me(3,3) is 2,2'-((3",4"-dimethyl-[2,2':5',2":5",2"':5"',2""-quinquethiophene]-5,5""-diyl)bis(methanylylidene))dimalononitrile) | Active layer | 40 nm, 33 wt% $C_{60}$, 80°C substrate temperature | Synthon Chemical GmbH, Germany |
| $C_{60}$ (Buckminster fullerene) | Hole blocking layer | 15 nm | Lumtec, Taiwan |
| $C_{60}$:$W_2(hpp)_4$ ($W_2(hpp)_4$ is Tetrakis(1,3,4,6,7,8-hexahydro-2H-pyrimido[1,2-a]pyrimidinato-ditungsten(II))) | Electron transporting layer | 5 nm, 3 wt% $W_2(hpp)_4$ | Novaled AG, Germany |
| ITO (Indium-tin-oxide) | Cathode | 90 nm | Thin Film Devices Inc., USA |
| Glass substrate | Substrate | $10^6$ nm | Thin Film Devices Inc., USA |

## 6. Sub-nanometer measurements

In the main manuscript, we demonstrated the wavelength sensor resolving excitation wavelength steps of 1 nm. Here, we present two measurements where the monochromator stepsize is decreased down to 0.1 nm. This is a resolution, where mechanical, thermal, and technical inaccuracies of the whole setup, i.e. the monochromator unit, may outcompete the resolution of our novel sensor. We, therefore, do not claim the 0.1 nm measurement to purely reflect the physics of our film.

Figure S6a and S6b present the afterglow intensity of PMMA:BP-2TA:QDs and the absorbance profile of PMMA:BP-2TA over excitation wavelength. Both profiles are similar. This is reasonable, as the absorbance of BP-2TA is related to the share of triplet emission.

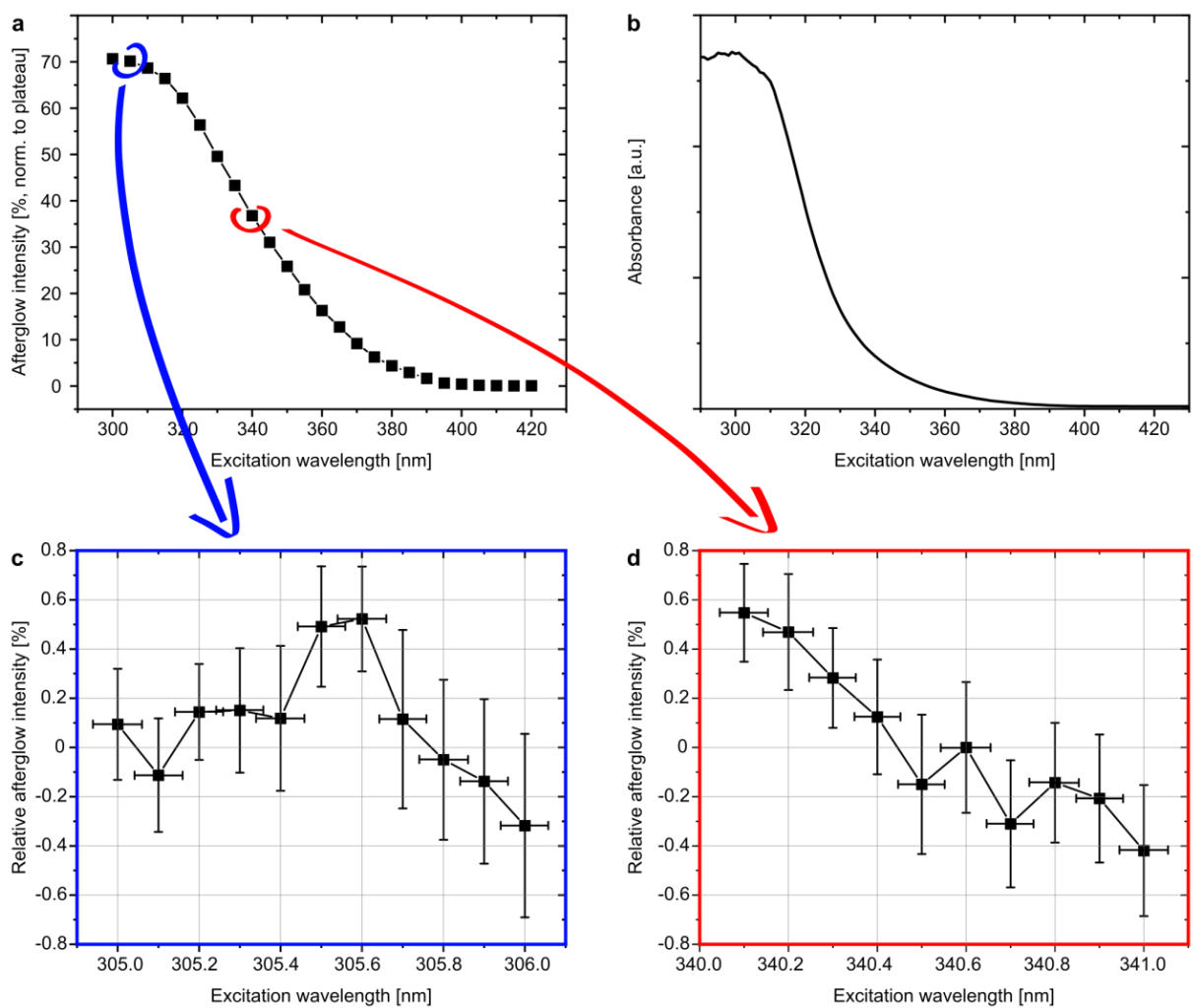

*Fig. S6. Relation between absorbance profile and tracking resolution. (a) Shows the afterglow intensity for PMMA:BP-2TA:QDs and (b) the absorbance profile for PMMA:BP-2TA. (c) and (d) show the zoom into regions at 305 nm and 340 nm. The wavelength reproducibility of the monochromator (Bentham MSH-300) is 0.05 nm and indicated as uncertainty. Its wavelength accuracy is 0.2 nm (Quantum Design, MSH-300 specification sheet).*

Figures S6c and S6d are taken at two different gradients of the afterglow intensity profile. At 305 nm, the gradient is small and, thus, the wavelength resolution is lower. Taking the standard deviations at each wavelength step into account (10 averages), it is not possible to resolve wavelength steps better than 1 nm. By contrast, the measurement around 340 nm, where the afterglow intensity gradient is highest, clearly shows that a sub-nanometer resolution is possible.

Those measurements were performed using the same experimental settings as in the main manuscript. Adaptations, such as an increase in averages (here 10 for each wavelength) or increasing the excitation intensity can reduce the standard deviation of the results and further improve the resolution.

## 7. Excitation light intensity characteristics

Figure S7 presents the intensity characteristics of the xenon light source + shutter configuration and one electrically driven LED. While the former takes about 4 ms to drop from 90% to 10%, the latter drops by about 3 orders of magnitude within less than 1 ms. The intensity profiles are taken under the same experimental conditions as described in the main manuscript using no PL sample.

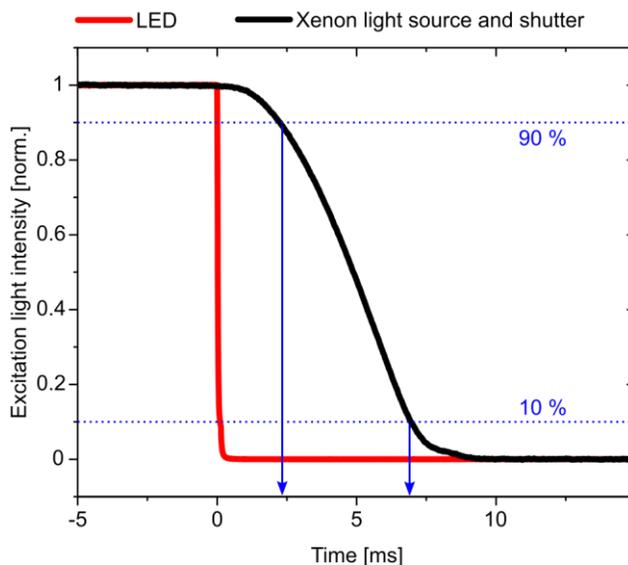

*Fig. S7. Temporal intensity characteristics of the excitation light sources. The 90%-to-10% intensity drop of the configuration using the xenon light source + shutter is indicated in blue lines.*

Figure S8 presents the intensity profile of the xenon light source + monochromator combination used throughout our experiments. The profile follows the irradiance distribution of the xenon white-light source and is superimposed by the performance of the gratings used for respective wavelength bands. The jump in intensity at 400 nm is due to a grating change that is always conducted at this point.

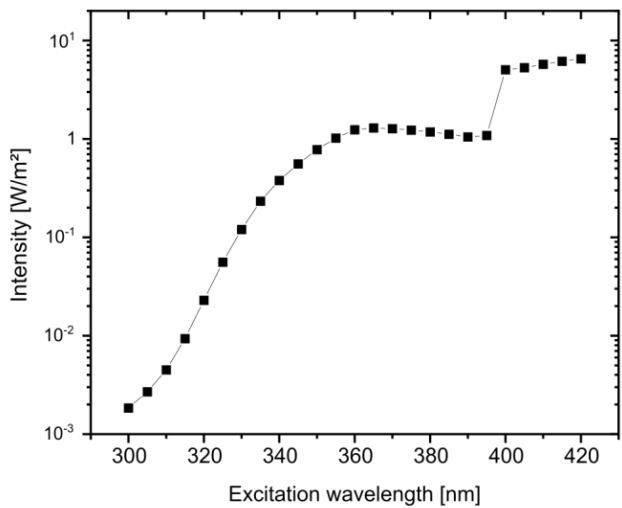

*Fig. S8. Intensity profile of the white light source (Xenon) in combination with the monochromator.*

## 8. Wavelength tracking of heating-induced LED emission shift

If driven constantly at the maximum current of 700 mA, the 365 nm Thorlabs LED features a peak wavelength shift of about 0.6 nm within 18 minutes (Fig. S9c). A spectral peak shift is indicated in the specification sheet due to a case temperature increase while operating (*5*). The off-cycle PL response of the BP-2TA:QD film shifts accordingly and can be correlated to the excitation peak wavelength of the LED (Figs. S9a and b). Within this limited wavelength range, the drift can be approximated by a linear regression which is then used as a "look-up table", used in the respective LED heating experiment in the main manuscript (Fig. S9d).

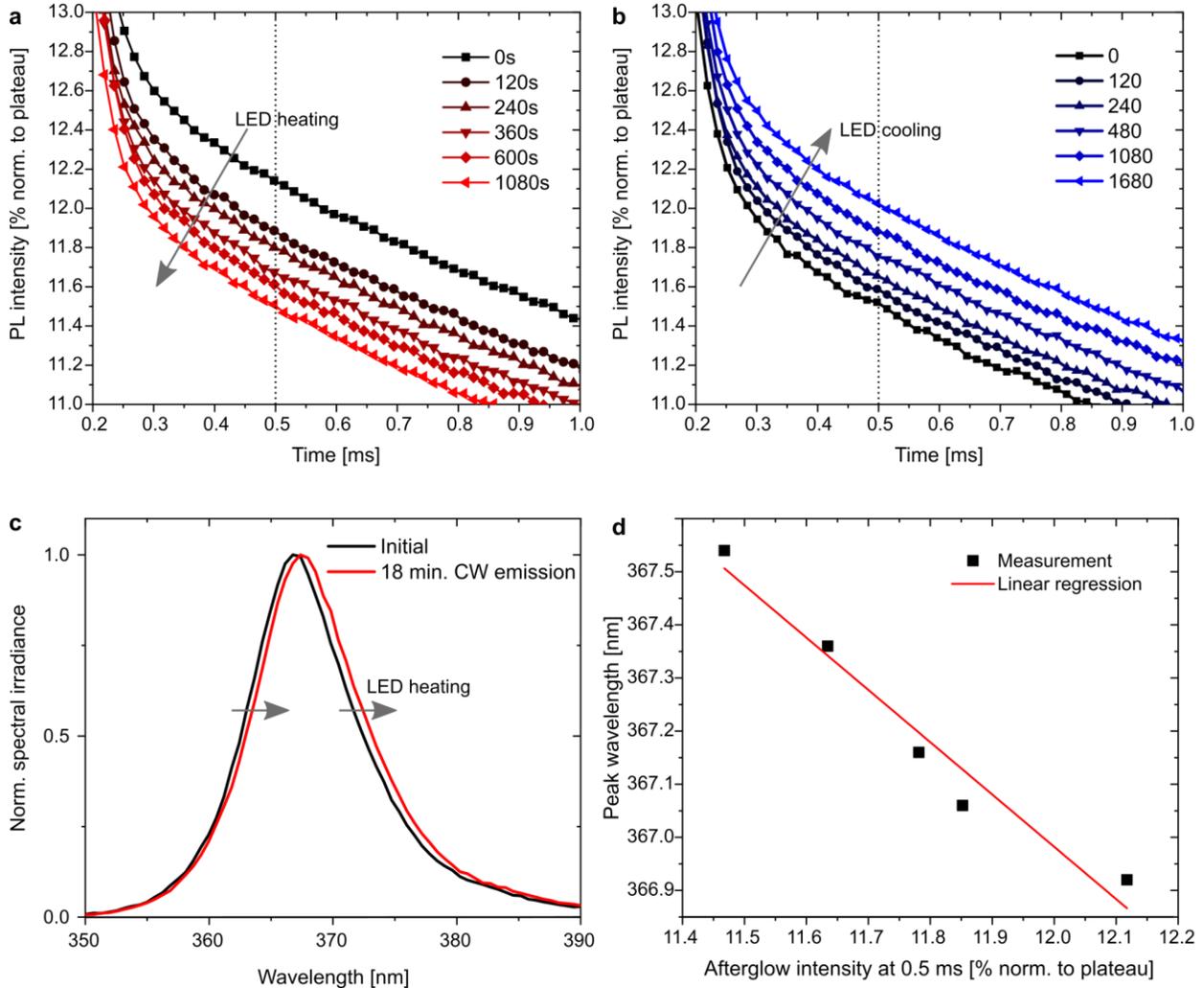

*Fig. S9. Wavelength tracking of LED heating. (a) Change in off-cycle PL transients while the LED is continuously operated at 700 mA. (b) Change in off-cycle PL transients while the LED rests in between the measurements. The readout point is indicated by a vertical, dotted line. (c) LED emission shift after running for 18 minutes. (d) Linear regression that correlates afterglow intensity at 0.5 ms to excitation peak wavelength.*

Figure 5 in the main manuscript shows that even after 30 minutes of rest the initial PL response of the film does not exactly match the initial condition. In further measurements, we saw that we always seem to have an overlaid trend distorting the measured PL data that looks like the heating feature of the controlling units and a slight shift in the trigger signal. This very delicate measurement, therefore, does not purely reflect the PL characteristics of the film but is also influenced by the thermal trajectories of the other experimental equipment. The systematic error of the measurement is hence larger than the statistical standard deviation of the measurements indicated in Fig. 5a in the main manuscript. The LED characteristics can, however, be identified very clearly.

### 9. Laser wavelength tracking

For laser wavelength tracking, a tunable dye laser is used, as specified in the Methods section. The laser uses Butyl-PBD (2-[1,1'-biphenyl]-4-yl-5-[4-(1,1-dimethylethyl)phenyl]-1,3,4-oxadiazole) dissolved in toluene at a concentration of $4 \cdot 10^{-3}$ molar to downconvert the 337 nm excitation laser. The converted emission wavelength can be tuned around 370 nm. In the main manuscript, a back-and-forth scan between 370 nm and 375 nm is presented. Two PDA100 Thorlabs photodetectors, each equipped with a 400 nm long-pass filter are used in this experiment. One detects the prompt PL burst of the active film (low amplification setting to not run into saturation and high temporal resolution). The other detects the very weak long afterglow (high amplification setting, low temporal resolution). The afterglow intensity is then normalized to the integrated prompt PL signal, which thus gives arbitrary units that are not comparable to the afterglow intensities measured in the other experiments. The PDA100 cannot truly resolve the prompt PL pulse but gives a voltage response that is proportional to its intensity.

Setting the laser to certain emission wavelengths and measuring the PL response of the film calibrates the system (cf. Fig. S10) and enables the back-and-forth sweep presented in the main manuscript. Despite taking 10 averages at every tuning step, the standard deviation of every measurement point is large, as the afterglow intensity is very weak due to the short excitation pulse.

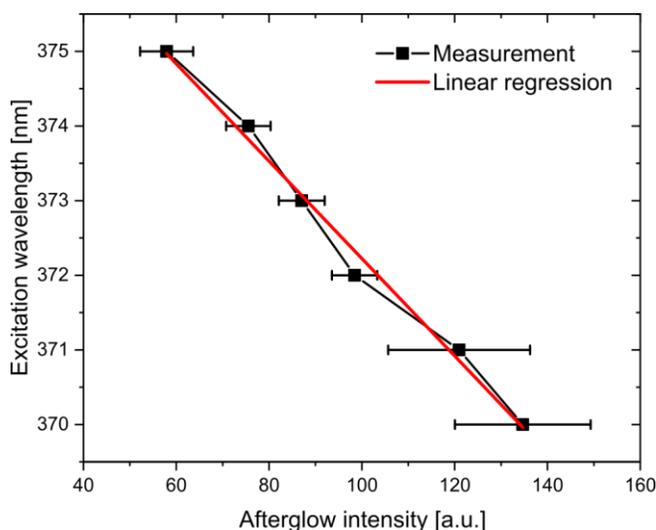

*Fig. S60. Laser wavelength tracking. A calibration measurement creates again a "look-up table" that relates the afterglow intensity of the off-cycle PL signal to the excitation wavelength.*

### 10. Film thickness measurements

The PL films were prepared by drop-casting 200µl of the respective solution onto an inch-by-inch glass substrate. Their thickness was determined using a Dektak 150 (Veeco) profilometer. The BP-2TA:DCJTB film has a thickness of 20 µm and the BP-2TA:QD film of about 30 µm. The thickness varies by a couple of micrometers across the surface, as drop-casting does not lead to very smooth films.

## Supplementary References